# Simulation of an imaging system for internal contamination of lungs using MPA-MURA coded aperture collimator

ZHANG Ting [1], WANG Lei [1,*], NING Jing [3], LU Wei [1], Wang Xiaofei[1], Zhang Hai-wei [1], TUO Xian-guo[2]

[1]Chengdu University of Technology, Chengdu, 610059, China

[2]Sichuan University of Science & Engineering, Zigong, 643000, China

[3]Beijing Institute of Radiation Medicine, Beijing, 100850, China

**Abstract** The nuclides inhaled during nuclear accidents usually cause internal contamination of the lungs with low activity. Although a parallel-hole imaging system, which is widely used in medical gamma cameras, has a high resolution and good image quality, owing to its extremely low detection efficiency, it remains difficult to obtain images of inhaled lung contamination. In this study, the Monte Carlo method was used to study the internal lung contamination imaging using the MPA-MURA coded-aperture collimator. The imaging system consisted of an adult male lung model, with a mosaicked, pattern-centered, and anti-symmetric MURA coded-aperture collimator model and a CsI(Tl) detector model. The MLEM decoding algorithm was used to reconstruct the internal contamination image, and the complementary imaging method was used to reduce the number of artifacts. The full width at half maximum of the I-131 point source image reconstructed by the mosaicked, pattern-centered, and anti-symmetric Modified uniformly redundant array (MPA-MURA) coded-aperture imaging reached 2.51 mm, and the signal-to-noise ratio of the simplified respiratory tract source (I-131) image reconstructed through MPA-MURA coded-aperture imaging was 3.98 dB. Although the spatial resolution of MPA-MURA coded aperture imaging is not as good as that of parallel-hole imaging, the detection efficiency of PMA-MURA coded-aperture imaging is two orders of magnitude higher than that of parallel hole collimator imaging. Considering the low activity level of internal lung contamination caused by nuclear accidents, PMA-MURA coded-aperture imaging has significant potential for the development of lung contamination imaging.

## 1. INTRODUCTION

There are several artificial radionuclides released into the public environment during the Three Mile Island, Chernobyl, and Fukushima accidents. The main nuclides released included Mo-99 (140.5 keV), Ce-141 (145 keV), Ce-144 (133 keV), I-131 (365 keV), Cs-137 (662 keV), Cs-134 (604 keV, 795 keV), Zr-95 (724.1 keV, 756.7 keV), Mn-54 (834.85 keV), and Co-60 (1173 keV, 1332 keV), [1–3] and enter the human body through breathing, drinking, eating, and open wounds. Therefore, it is imperative to urgently carry out the screening of internal radioactive contamination as early as possible after a nuclear accident [4–7].

Although it is assumed that the radionuclides that enter the body are uniformly distributed in the organs by measuring the internal contamination activity, efficiency calibration, and dose calculation in the internal contamination assessments, this is not the case. Owing to the varied physical and chemical characteristics of different nuclides and individual metabolism, the distribution of different radionuclides in different organs changes with metabolism and time. The heterogeneity, time variability, and individuality of the distribution of radionuclides in different organs are considered key points that cause errors in individual internal radiation dose assessments. Therefore, it is extremely meaningful to obtain the distribution of radionuclides in the body based on imaging measurements, which assist in acquiring a more accurate internal radiation dose.

A medical gamma camera with a parallel-hole collimator usually has a high spatial resolution, but an extremely low detection efficiency. The activities of nuclides introduced into patients during medical diagnosis generally reach the order of $10^6$Bq. However, the activity level of contamination in individuals caused by a nuclear accident is usually lower than $10^5$Bq [8]. Therefore, it becomes difficult for medical gamma cameras to acquire high-quality images of low-activity of internal contamination within a short time.

A coded-aperture collimator might be a good choice for obtaining internal contamination of the lungs through rapid imaging. Since coding imaging technology was first proposed in 1961, it has evolved from random arrays (RAs) into non-redundant arrays (NRAs), uniform redundant arrays (URAs), and modified uniform redundant arrays (MURA) [9–12]. The opening rate of a MURA can reach 50%, and the transmittance and detection efficiency of a MURA coded-aperture collimator are much higher than those of the parallel-hole imaging. To investigate the internal contamination imaging using MURA coded-aperture imaging, the Monte Carlo method was used to simulate the images of three nuclides with four distributions deposited in the lung model by applying an MPA-MURA coded-aperture collimator. The results show that, although the spatial resolution of the MPA-MURA coding aperture imaging is not as good as the parallel-hole collimator imaging, its detection efficiency is two orders of magnitude better than the latter. Thus, MPA-MURA coded-aperture imaging assists us in improving the estimation of the internal contamination.

## 2. MODEL ESTABLISHMENT

This paper presents the design of an internal lung MPA-MURA imaging system based on Geant4, including a lung model of an adult male, an MPA-MURA coding hole collimator model, and a detector model.

### 2.1 Lung model of adult male

The MIRD phantom is a stylized computational phantom developed in the 1960s. It is represented mathematically and can be coupled with a Monte Carlo radiation transport computer code to track the radiation interactions and energy deposition in the body. Among the 140 human models available internationally, the MIRD model is the primary programmatic model for evaluating the exposure calculations in the human body. It simplifies the organs into a series of basic three-dimensional shapes and spatial combinations. The geometric features of these organs are described using mathematical formulas, and finally, the individual organs are combined into a human body model.

The tissues and organ models of the chest, including both lungs and the heart, thymus, spine, ribs, muscle, and soft tissues, from an adult male MIRD were cropped in Geant4. The size of the lungs of the adult male MIRD phantom was 290 mm × 240 mm (Fig. 3 (c)). Therefore, a 301 mm × 301 mm field of view (FoV) of the imaging system was designed to completely image the lungs.

#### 2.1.1 MPA-MURA coding hole collimator model

The coded aperture collimator, a key device for imaging, is responsible for encoding the incident gamma rays such that the detector can obtain a coded image of the object. The coded aperture collimator model used in this study uses a mosaicked, pattern-centered, and anti-symmetric MURA (MPA-MURA).

#### 2.1.2 Matrix function of MPA-MURA

It is assumed that both the number of rows and columns of the MURA are P (P is a prime number), and MURA is represented by matrix A, as shown in (1) [13]. Each element of $A$ can be a 1 or zero, where 1 represents an opaque block and zero represents a hole.

$$(\mathrm{A_{ij}})_{\mathrm{p\times p}} = \begin{cases} 0 & if \quad i = 0 \\ 1 & if \quad j = 0, i \neq 0 \\ 1 & if \quad c_p(i)c_p(j) = 1 \\ 0 & otherwise \end{cases} \tag{1}$$

where $c_{\mathrm{p}}(\mathrm{i})$, $\mathrm{c_p}(\mathrm{j})$ satisfies the following:

$$c_p(i) = \begin{cases} 1 & if \ \exists x, 0 < x < p, i = mod_r(x^2) \\ -1 & otherwise \end{cases}$$

To facilitate decoding of the coded images, MURA matrix A was transformed into a center-symmetric matrix through the following steps.

In the first step, matrix A was divided into seven sub-regions, R, S, T, U, V, W, X, Y, and Z, as shown in (2) and (3).

$$(A)_{p\times p} = \begin{bmatrix} A_{00} & A_{01} & A_{02} & \cdots & \cdots & A_{0,\frac{P-1}{2}} & A_{0,\frac{P+1}{2}} & \cdots & \cdots & A_{0,P-1} \\ A_{10} & A_{11} & A_{12} & \cdots & \cdots & A_{1,\frac{P-1}{2}} & A_{1,\frac{P+1}{2}} & \cdots & \cdots & A_{1,P-1} \\ A_{20} & A_{21} & A_{22} & \cdots & \cdots & A_{2,\frac{P-1}{2}} & A_{2,\frac{P+1}{2}} & \cdots & \cdots & A_{2,P-1} \\ \vdots & \vdots & \vdots & \cdots & \cdots & \vdots & \vdots & \cdots & \cdots & \vdots \\ \vdots & \vdots & \vdots & \cdots & \cdots & \vdots & \vdots & \cdots & \cdots & \vdots \\ A_{\frac{P-1}{2},0} & A_{\frac{P-1}{2},1} & A_{\frac{P-1}{2},2} & \cdots & \cdots & A_{\frac{P-1}{2},\frac{P-1}{2}} & A_{\frac{P-1}{2},\frac{P+1}{2}} & \cdots & \cdots & A_{\frac{P-1}{2},P-1} \\ A_{\frac{P+1}{2},0} & A_{\frac{P+1}{2},1} & A_{\frac{P+1}{2},2} & \cdots & \cdots & A_{\frac{P+1}{2},\frac{P-1}{2}} & A_{\frac{P+1}{2},\frac{P+1}{2}} & \cdots & \cdots & A_{\frac{P+1}{2},P-1} \\ \vdots & \vdots & \vdots & \cdots & \cdots & \vdots & \vdots & \cdots & \cdots & \vdots \\ \vdots & \vdots & \vdots & \cdots & \cdots & \vdots & \vdots & \cdots & \cdots & \vdots \\ A_{P-1,0} & A_{P-1,1} & A_{P-1,2} & \cdots & \cdots & A_{P-1,\frac{P-1}{2}} & A_{P-1,\frac{P+1}{2}} & \cdots & \cdots & A_{P-1,P-1} \end{bmatrix} \tag{2}$$

$$
\begin{cases}
R_{i,j} = A_{i,j} & i = 0;\ j = 0,1,\dots\dots,\dfrac{p-3}{2} \\
T_{i,j} = A_{i,j+\frac{p+1}{2}} & i = 0;\ j = 0,1,\dots\dots,\dfrac{p-3}{2} \\
S = A_{0,\frac{p-1}{2}} & \\
U_{i,j} = A_{i+1,j} & i = 0,1,2,\dots\dots,\dfrac{p-3}{2};\ j = 0 \\
V_{i,j} = A_{i+\frac{p+1}{2},j} & i = 0,1,2,\dots\dots,\dfrac{p-3}{2};\ j = 0 \\
W_{i,j} = A_{i+1,j+1} & i = 0,1,\dots\dots,\dfrac{p-3}{2};\ j = 0,1,\dots\dots,\dfrac{p-3}{2} \\
X_{i,j} = A_{i+1,j+\frac{p+1}{2}} & i = 0,1,\dots\dots,\dfrac{p-3}{2};\ j = 0,1,\dots\dots,\dfrac{p-3}{2} \\
Y_{i,j} = A_{i+\frac{p+1}{2},j+1} & i = 0,1,\dots\dots,\dfrac{p-3}{2};\ j = 0,1,\dots\dots,\dfrac{p-3}{2} \\
Z_{i,j} = A_{i+\frac{p+1}{2},j+\frac{p+1}{2}} & i = 0,1,\dots\dots,\dfrac{p-3}{2};\ j = 0,1,\dots\dots,\dfrac{p-3}{2}
\end{cases}
\tag{3}
$$

In the second step, the *W*, *X*, *Y*, and *Z* sub-regions were transformed into *E*, *F*, *G*, and *H*, respectively, by rotating 180°, as shown in (4).

$$
\begin{cases}
E_{kt} = W_{\left(\frac{P-1}{2}-1-k\right)\left(\frac{P-1}{2}-1-t\right)} \\
F_{kt} = X_{\left(\frac{P-1}{2}-1-k\right)\left(\frac{P-1}{2}-1-t\right)} \\
G_{kt} = Y_{\left(\frac{P-1}{2}-1-k\right)\left(\frac{P-1}{2}-1-t\right)} \\
H_{kt} = Z_{\left(\frac{P-1}{2}-1-k\right)\left(\frac{P-1}{2}-1-t\right)}
\end{cases}
\tag{4}
$$

where $\mathrm{k} = 0, 1, \dots\dots \frac{p-1}{2} - 1, \mathrm{t} = 0, 1, \dots\dots \frac{p-1}{2} - 1$.

A was transformed into B as shown in (5).

$(B)_{p\times p}=$

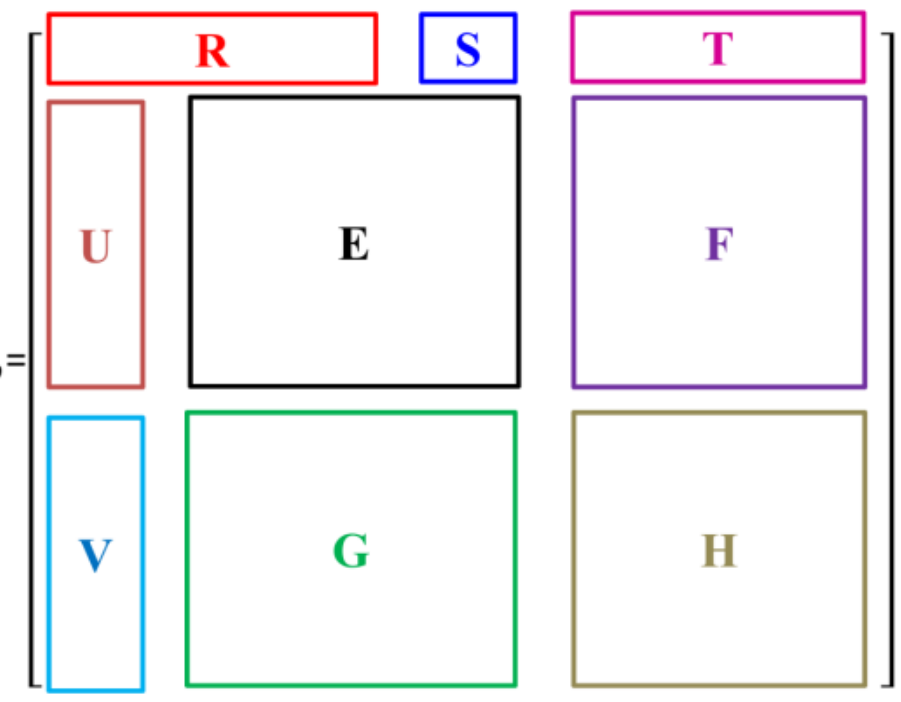

(5)

In the third step, matrix B is transformed into matrix C by moving *R*, *S*, *T*, *U*, and *V* toward the center and *E*, *F*, *G*, and *H* toward the sides, as shown in (6). Matrix C was a pattern-centered and anti-symmetric MURA.

$(C)_{p\times p}=$

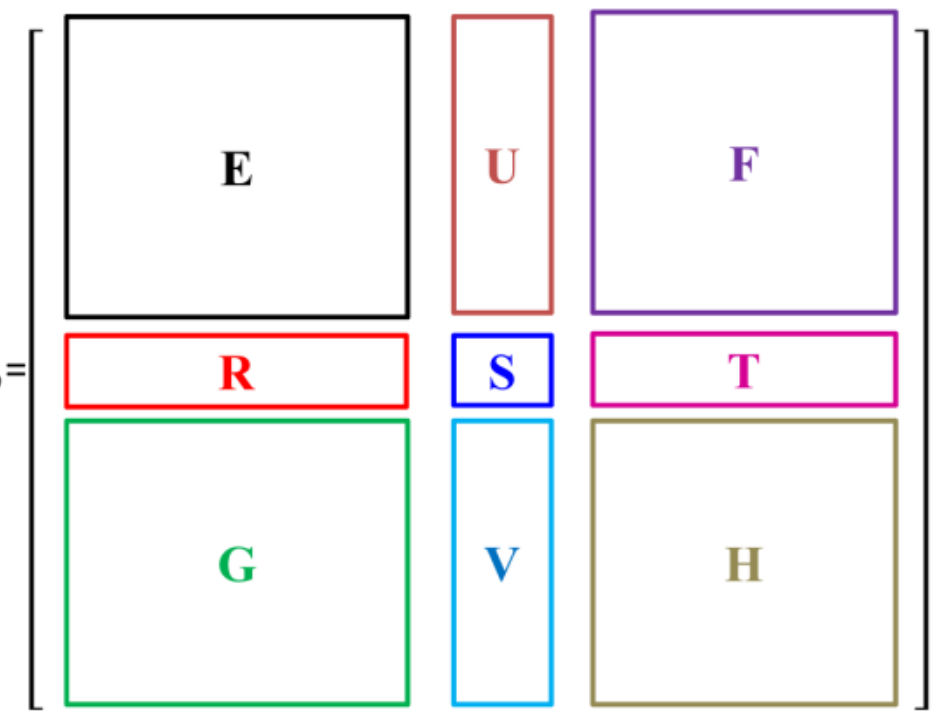

(6)

In the fourth step, the C matrix was extended to form a mosaicked, pattern-centered, and anti-symmetric MURA ($D_{q\times q}(q = 2p - 1)$ )[14]. First, matrix C was split into left and right parts along the central column. The left and the right parts were folded toward the right and left sides of matrix C, respectively. Second, matrix C was split into the upper and lower parts along the middle row, and the upper and lower parts were folded toward the lower and upper sides of matrix C. Finally, matrix C was divided into four parts along the middle row and middle column, and were moved to the respective four corners of matrix C. Matrix C and the eight extended parts created in the above three steps were used to create a new matrix D. Matrix D was a mosaicked, pattern-centered, and anti-symmetric MURA, as shown in (7).

$$(D)_{q\times q}=\begin{bmatrix} H & G & V & H & G \\ F & E & U & F & E \\ T & R & S & T & R \\ H & G & V & H & G \\ F & E & U & F & E \end{bmatrix} \quad (7)$$

**2.1.3 Specification of MPA-MURA**

The imaging geometry is shown in Fig. 1, where FoV is the field of view of the imaging system, O is the origin of the coordinates, and $D_1D_2$ is the length of the sides of the detector plane and is expressed as $d_d$. In addition, $a$ is the object distance, and $b$ is the focal distance. Moreover, $A_1A_2$ is the length of the sides of the PA-MURA coded-aperture plane, and $M_1M_2$ is the length of the sides of the MPA-MURA coded-aperture collimator, and is expressed as $d_m$.

Assuming that the number of rows of the PA-MURA coded-aperture plane is q and the number of rows of the MPA-MURA coded-aperture plane is $p$, according to the transformation of PA-MURA to MPA-MURA, we can obtain a $p$ value that is equal to $2q$-1. The object distance and focal distance of the imaging system were set to the same values, that is, $a = b$. To obtain the image of the full lungs, the FoV (F1F2) and M1M2 were both designed as 301 mm. The length of the sides of the hole in the MPA-MURA coded-aperture collimator was set to 1 mm. Therefore, the number of rows of MPA-MURA was 301, and the number of rows of the PA-MURA was 151. The area of the MPA-MURA coded-aperture collimator was 301 mm × 301 mm.

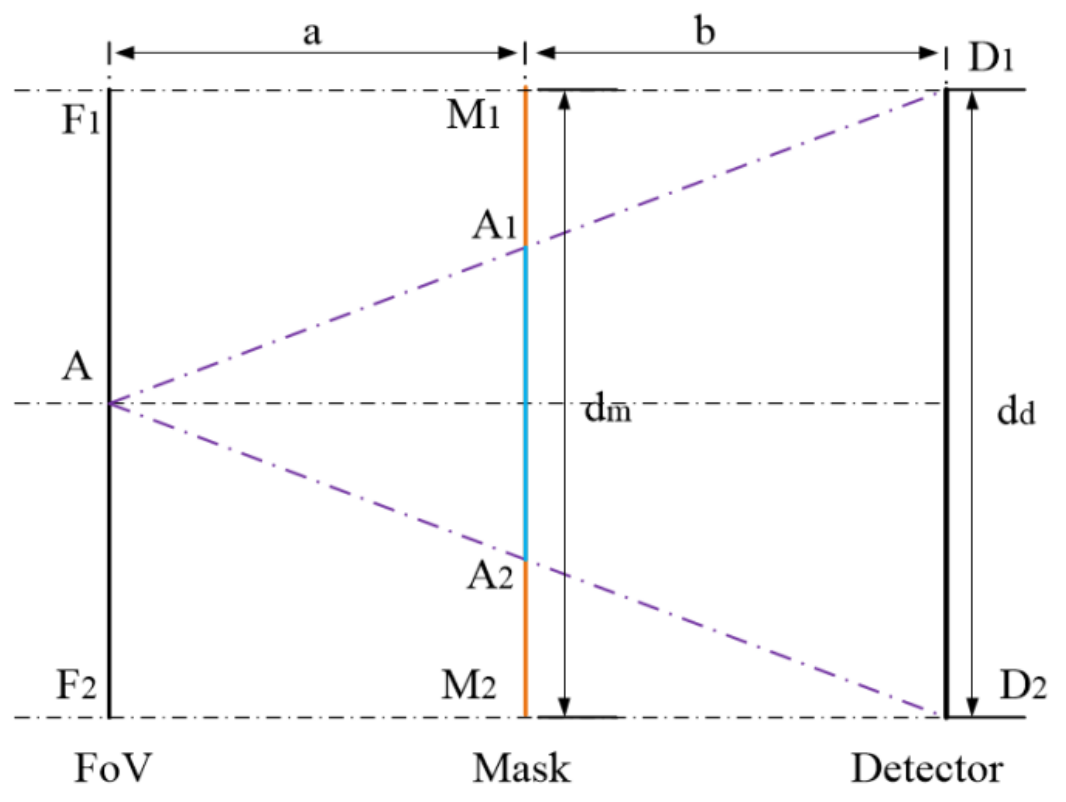

Fig. 1 Geometry of imaging system.

### 2.1.4 Materials of MPA-MURA

Tungsten steel was chosen as the material for the coded hole collimators because of its high melting point, high modulus of elasticity, extremely low coefficient of thermal expansion, good linear attenuation characteristics, and ease of processing and formation. The coded hole collimator in this study is shown in Fig. 3 (a).

## 2.2 Detector model

Nuclear radiation detectors are the most important part of an imaging system. The most widely used detector material at this stage is the scintillator crystal. A CsI(Tl) scintillator crystal was employed to create a position-sensitive detector model. Energy broadening and an energy resolution were not considered in the simulations in this study. Only the number of photons, which deposited energy in the crystal, was recorded and used to form the coded image. The size of the pixel matrix of the position-sensitive detector model was $N_d \times N_d$, where $N_d = \alpha \times q$[15]. Here, $q$ is the number of rows of the MPA-MURA coded-aperture collimator model, and α is the sampling factor of the detection system. In the imaging system, $\alpha = 1$ and $q = N_d = 301$. The size of the pixel of the position-sensitive detector model was set to 1 mm × 1 mm [16], and the thickness of the CsI(Tl) crystal was set to 38.1 mm, as shown in Fig. 3 (b). The detector grid has 301 × 301 = 90,601 small regions, numbered from 1 to 90,601, and all blank areas other than this detector array are numbered as zero. The G4GetCopyNumber step processing method is used to record the number of particles in different grid areas of the detector. When a particle enters the detector, if the spatial serial number changes from zero to the detector number, the detector count plus 1 is used, and the final count will be the detector count for each grid area.

## 2.3 Thickness of MPA-MURA and source-detector distance

When the gamma rays pass through the coded-aperture collimator, some of them directly reach the detector through the holes of the collimator, and some rays can interact with the collimator and scatter. The scattered rays might enter the detector by deviating from the original direction, which then generates noise in the coded image and enhances the artifacts. To optimize the imaging quality and detection efficiency, the I-131 point source was used to optimize the imaging geometry and the thickness of the coded-aperture collimator. The effect of the source-detector distance (Z) and the thickness of the MPA-MURA coded-aperture collimator (t) on the imaging quality and detection efficiency were calculated using a Monte Carlo simulation. The simulation satisfies the following conditions: (1) the I-131 point source was placed in the lungs, (2) the object distance and the focal distance were set to the same value, whereas the source-detector distance (Z) varied from 40 to 240 mm, and (3) the lung model shown in Fig. 3(c), the coded-aperture model shown in Fig. 3(a), and the detector model shown in Fig. 3(b) were used.

According to Fig. 2(a), when the thickness of the coding hole collimator was 4 mm, the figure of merit (FOM) [30] achieved the local minimum at each source-detector distance (Z), varying from 40 to 240 mm. Therefore, the thickness of the coding hole collimator was set to 4 mm in the following imaging simulation.

As shown in Figs. 2(a) and 2(b), while the source-detector distance (*Z*) increases, the imaging quality (FOM) improves, but the detection efficiency gradually decreases. To balance between the detection efficiency and the imaging quality, the source-detector distance (*Z*) was selected to be 160 mm, and both the object distance (*a*) and the focal distance (*b*) were set to 80 mm, as shown in Fig. 3(c).

The mathematical definition of FoM is as follows [30]:

$$\text{FoM} = \sqrt{\frac{\iint\left[(x-\mu_x)^2+\left(y-\mu_y\right)^2\right]\times\hat{O}(x,y)dxdy}{\iint\hat{O}(x,y)dxdy}} \tag{8}$$

with

$$\mu_x = \frac{\iint x\times\hat{O}(x,y)dxdy}{\iint\hat{O}(x,y)dxdy} \qquad \mu_y = \frac{\iint y\times\hat{O}(x,y)dxdy}{\iint\hat{O}(x,y)dxdy}, \tag{9}$$

where $\widehat{O}(x,y)$ is the brightness of the reconstructed image of a point source.

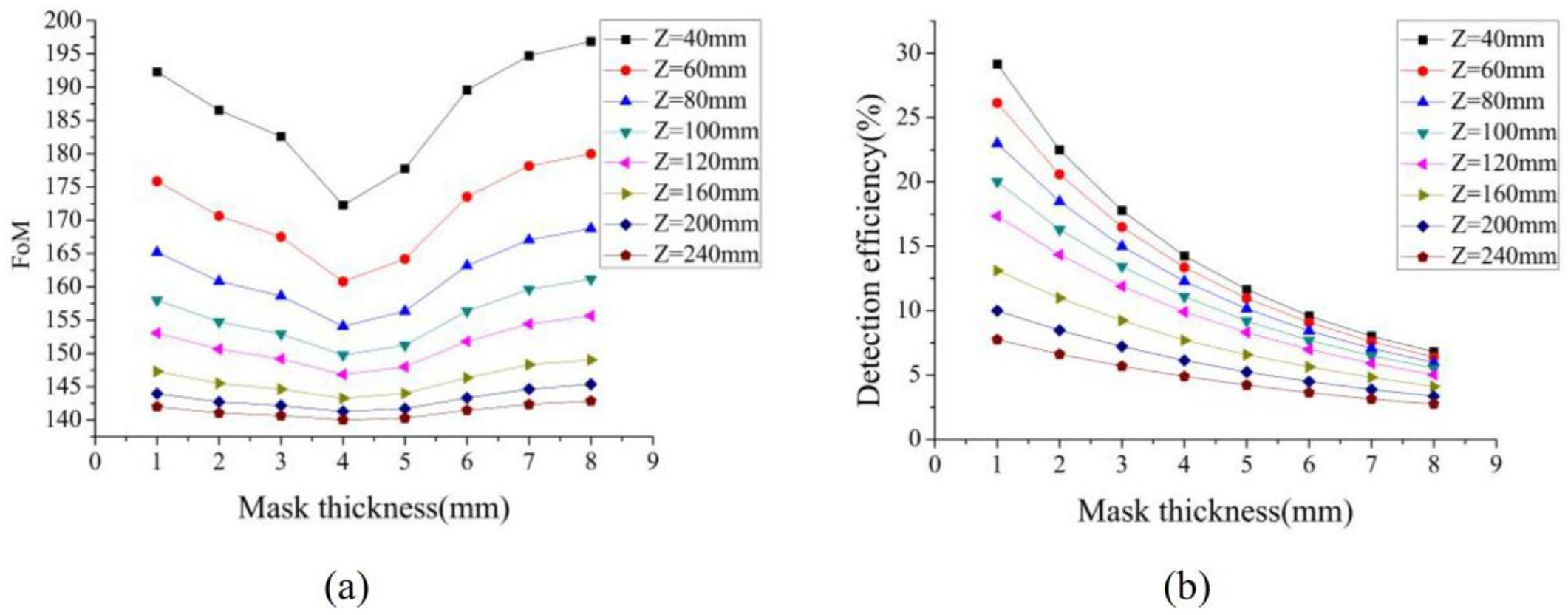

Fig. 2 I-131 point source reconstruction of image parameters. (a) The FOM as a function of mask thickness at different object-detector distances. (b) The detection efficiency as a function of mask thickness at different object-detector distances.

## 2.4 Imaging system

The internal lung contamination imaging system designed in this project is shown in Fig. 3(c), and the center of the lung model, the coded-aperture collimator model, and the detector model are shown on the axis. The lung model was cropped from an adult male MIRD phantom. The coded-aperture collimator model attained a 4-mm thick mosaicked, pattern-centered, and anti-symmetric MURA with a size of 301 × 301, including a pixel size of 1 mm × 1 mm, and used tungsten steel as the material.

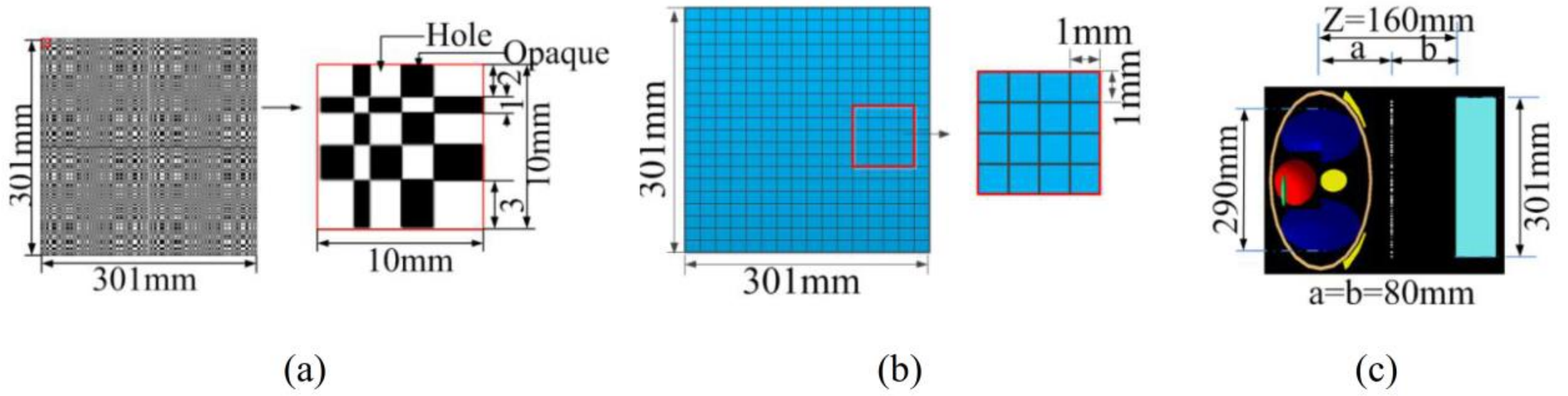

Fig. 3 (a) 301 × 301 mosaicked, pattern-centered, and anti-symmetric MURA coded-aperture collimator model.

(b) Position-sensitive detector model based on the CsI(Tl) scintillator crystal. (c) Model of the imaging system for internal lung contamination.

## 3. MAXIMUM LIKELIHOOD-EXPECTATION MAXIMIZATION

The decoding algorithm directly affects the quality of the reconstructed image. The iterative algorithm is a widely used decoding algorithm for image reconstruction. In this study, the maximum likelihood expectation maximization (MLEM) iterative algorithm was employed to reconstruct the radiation image. In the MLEM iterative algorithm, the initial estimated value of the image was set as 1, where $X^0(x,y) \equiv 1$. First, the current estimated image was projected as $X^k(x,y)$ to the detector, and the coded-aperture array was used to obtain an estimated coded image. Second, the back-project ratio of the coded image $Y(x,y)$ obtained by the detector in the imaging simulation experiment to the estimated image $X^k(x,y)$ used to estimate the space was obtained as a correction factor. Finally, the correction factor was used to correct the estimated image $X^k(x,y)$. This process of projection and back-projection was repeated until the iteration stop condition $\mathrm{Max}_{(x,y)}(|\frac{X^{k+1}(x,y)}{X^k(x,y)} - 1|) \leq 0.005$ was satisfied.

The MLEM algorithm can be expressed as (8) [17]:

$$X^{k+1}(x,y) = X^k(x,y) \times [\frac{Y(x,y)}{X^k(x,y) * h(x,y)} \otimes h(x,y)] \quad (10)$$

where $X^k(x,y)$ is the estimated image after the kth iteration, $Y$ (x, y) is the encoded image obtained by the detector, $h$ (x, y) is the encoded matrix function, and ⊗ is a period correlation operation.

## 4. COMPLEMENTARY IMAGING

Under near-field imaging conditions, it is difficult for a coded-hole collimator to provide an ideal point spread function. The different incident angles will change the projection intensity and then generate an intensity modulation to create artifacts [18–21]. Complementary coded-aperture imaging was used to reduce noise and artifacts [22]. As the coded-hole collimator designed in this study was mosaicked, pattern-centered, and antisymmetric, a positive coded image was used and then rotated by 90° around the center to obtain a reverse coded image. The positive and reverse coded images were decoded by the MLEM algorithm to acquire a positive image and a reverse image of the source, respectively. Finally, the reconstructed image was obtained from the source by subtracting the reverse image from the positive image, as shown in Fig. 4.

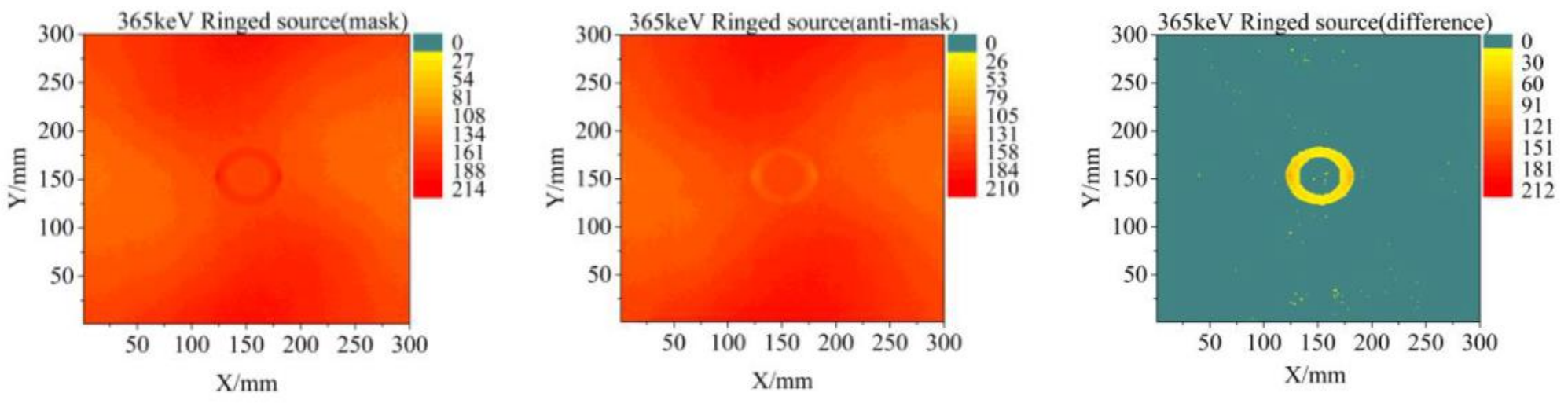

(a) (b) (c)

Fig. 4. A ring source rebuilt using a complementary imaging method: (a) Ring source image obtained when the coded-aperture collimator was at zero degrees, (b) the ring source image obtained when the coded-aperture was rotated by 90º around the center, and (c) the ring source image equivalent to (a) and (b).

As can be seen in Fig. 4 (a), the counts are higher in the circular radiogenic region at the center of the image, the artifact counts are lower on the left and right sides, and the artifact counts on the upper and lower sides are similar to those in the circular radiogenic region at the center. Fig. 4 (b) shows higher artifact counts on the upper and lower sides. The circular radioactive source region counts at the center of the image are lower, and the left- and right-side artifact counts do not differ much from the circular radioactive source region counts at the center. The presence of background noise and artifact counts around the circular radioactive source in (a) does not differ significantly from those in (b). Subtracting (a) and (b) yields (c), where the noise and artifacts are mostly eliminated. This shows that complementary imaging can effectively reduce the near-field artifacts in the reconstructed images of radioactive sources.

## 5. GEANT4 SIMULATION

The accuracy and usefulness of Geant4 simulations have been demonstrated in a number of cases. [23–29]

In the Geant4 software, the simulation process can be simplified into a few steps.

The first step is to build the models. The models in G4 are composed of geometric blocks of different sizes, the largest of which we have defined in the software as "World Volume." We define different elements by defining the atomic weight, proton number, and density, and then different elements to define the material for the detector and the coded hole collimator. The parameter settings of the system model are shown in Figure 3. To compare the MPA-MURA coded aperture imaging and the parallel hole imaging, the coded-aperture collimator was replaced with a parallel hole collimator under the same imaging conditions. The material of the parallel hole collimator used was tungsten steel, with a thickness of 1 mm, a hole interval of 0.5 mm, and collimator specifications of 340 mm × 340 mm × 80 mm. The resolutions of both the parallel hole collimator and the MPA-MURA coded aperture collimator were set to 2 mm.

The second step is to define the particles needed and the corresponding physical reaction processes. The first step is to define the types of particles needed to simulate the process by inheriting the seven major classes of particles that come with G4 through known knowledge of nuclear physics. We then need to set up the various reactions that the particles may have in the detector and add the corresponding physical processes to them. We can also use the standard physical process module that comes with G4, which allows customizing the physical reactions. Because this simulation is only concerned with imaging and does not involve the calculation of dosimetric parameters such as the absorption dose, G4EmStandardPhysics is used, and is included in the G4 Physics List Library (which is made up of standard physics processes and low-energy physics). The case is one of the main parts of the reference physics process. Standard electromagnetic physics procedures are used in many G4 examples, G4 tests, and G4 validation studies, where the default cut-off threshold for γ-rays, electrons, and positrons is set to 1 mm. The physical procedures of the Monte Carlo simulation program in this study mainly consider electromagnetic interactions. The electromagnetic interactions of $\gamma$ -rays and charged particles with the detector include photoelectric effects, Compton scattering, electron pair effects, ionization, toughened radiation, and multiple scattering.

The third step is to use the tracer processing classes of G4 to process the nuclear data. The G4 program processes nuclear data by using the basic classes of Stepping, Event, Tracking, and Run. To simulate the overall operation of the program, Run is used for control. In addition, Event is used to control each particle emission to be absorbed by the detector, and each time a particle undergoes a physical reaction such as excitation and ionization in the detector, the Step class uses the G4GetCopyNumber step processing method of G4 to record the number of particles in different grid regions of the detector. When the particles enter the detector, if the spatial sequence number changes from the original zero to the detector number, this is the count of the detector plus 1, and the final count is the count of each grid area of the detector. The final count is the count of each grid area of the detector, which is the information for the location of the radioactive source on the detector.

In the fourth step, different shapes of radioactive sources are generated, which are processed in the particle generation class by sampling the particle generation positions (PrimaryGeneratorAction.hh). By sampling the coordinates of the three directions (X, Y, Z) in a three-dimensional space according to a certain mathematical pattern, it is possible to make the radioactive source emit in a set manner, thus achieving a simulation of a radioactive source of different shapes. The same method can also be used to process the direction of the momentum of the radioactive source, decomposing the momentum of the radioactive source into three coordinates, sampling according to the angular distribution, and combining with the coordinate sampling. Different shapes and different emission directions of the radioactive source can be obtained. Using a particle gun (G4ParticleGun), the XYZ three planes are described using different mathematical formulas, and thus the coordinates of the particles are distributed in a three-dimensional space according to a regular pattern. The imaging of the four distributions of four nuclides (140keV Tc-99m, 365keV I-131, 662keV Cs-137, and 1332keV Co-60) in the lungs was simulated in this study. The four distributions are pointed, A-shaped, ringed, and a simplified respiratory tract. Geant4 was used to simulate the distributions of the four nuclides (140keV Tc-99m, 365keV I-131, 662keV Cs-137, and 1332keV Co-60) in the lungs. There were also four shaped distributions of the nuclides, i.e., pointed, A-shaped, ringed, and simplified respiratory tract, where the nuclides were evenly distributed on each shape. The width of the lines of the A-shaped source was 2 mm. The inner and outer diameters of the ring source were 20 and 30 mm, respectively. The width of the lines in the simplified respiratory tract varied from 2 to 10 mm. In the imaging simulation of each source, $1 \times 10^8$ photons were emitted.

## 6. RESULTS

### 6.1 FWHM of point source imaging

The localization accuracy of the reconstructed image is determined by the spatial geometric resolution of the system. Small localization errors in internal contamination imaging can have serious consequences, and thus its localization accuracy is extremely important. The resolution of the imaging system is determined based on the spatial geometric resolution of the collimator and the intrinsic resolution of the detector, which can occasionally be affected by the noise environment. In this paper, only the spatial geometric resolution of the collimator is discussed. This is defined by the full width at half maximum (FWHM) of the point spread function. An imaging system with the MPA-MURA coded-aperture collimator and the parallel hole collimator, respectively, was used to simulate the imaging of the point source of the four nuclides in the imaging geometry, as described in part V. In the MPA-MURA imaging system, the FWHM of the Tc-99m, Co-60, I-131, Cs-137, and Co-60 point-source reconstructed images were 2.48, 2.51, 2.64, and 3.08 mm, respectively. In the parallel hole imaging system, the FWHM of the Tc-99m, Co-60, I-131, Cs-137, and Co-60 point-source reconstructed images are 2.06, 2.13, 2.17, and 2.21 mm. The FWHM of the coded-aperture collimator imaging was 1.20- to 1.39-times that of the parallel hole collimator. The FWHM of the point source effect varies at different energies. In general, the higher the energy of the rays, the larger the FWHM. With the same thickness as the coded hole collimator, as the energy of the radiation source increases, the number of rays transmitted through the coded hole collimator increases, the FWHM of the point source effect increases, and the geometric resolution of the system increases. This is because high-energy rays are more likely to penetrate into the shield area of the collimator and increase the noise than the low-energy rays in both coded collimator imaging and parallel hole imaging.

The spatial geometric resolution set in the text was 2 mm. In the simulation, taking I-131 as an example, the geometric resolution of the MPA-MURA imaging system is 2.51 mm, with a relative error of 25.5% with a system design resolution of 2 mm, and the geometric resolution of the PB imaging system is 2.13 mm, with a relative error of 6.5% with a system design resolution of 2 mm. It can be seen that the geometric resolution of the PB parallel beam imaging system in the simulation is better than that of the MPA-MURA imaging system.

### 6.2 Simulation of internal contamination imaging

The internal contamination imaging was simulated in the imaging system using an MPA-MURA coded-aperture collimator and a parallel hole collimator, respectively. The three shape distributions of the internal contamination associated with Tc-99, I-131, Cs-137, and Co-60 in the lungs were imaged, and are presented in Fig. 5. The MPA-MURA encoded hole collimator was used to reconstruct the internal contamination distribution images using the MLEM algorithm after imaging. The noise and artifacts of the reconstructed distribution images were reduced using the complementary imaging method. Actually, the radioactive source *in vivo* has a three-dimensional distribution. Because we aim to use the plane detector to obtain the projection of the distribution, we simplified the sources with a two-dimensional distribution parallel to the detector plane during the simulations.

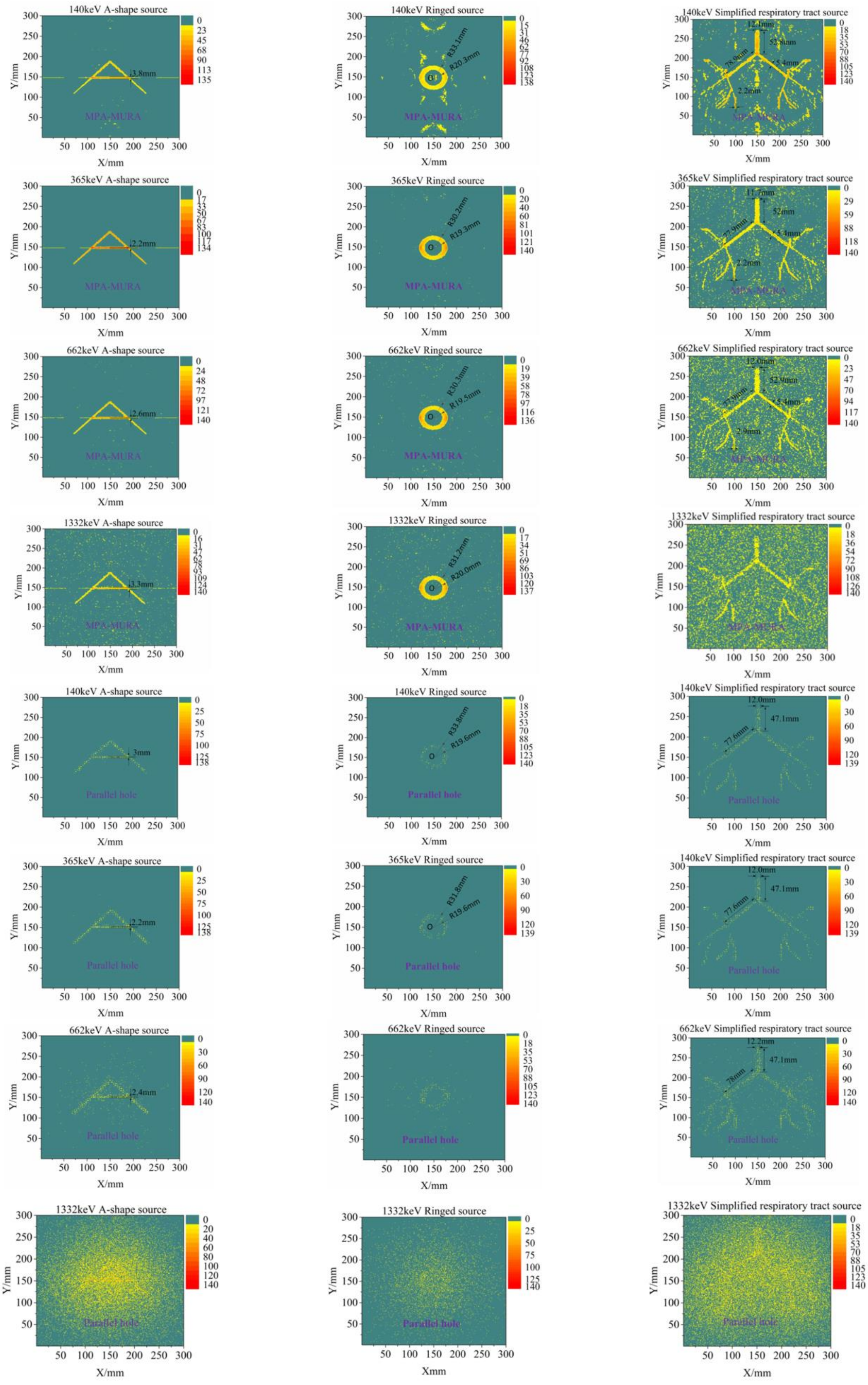

Fig. 5 Images of the internal contamination in the lungs acquired using the imaging system with a parallel

hole collimator and an MPA-MURA coded-aperture collimator.

As can be seen horizontally in Fig. 5, the quality of the reconstructed image decreases as the type of radioactive source becomes more complex. This is due to an increase in the number of radioactive sources and an increase in their interaction. As can be seen vertically in the figure, as the gamma-ray energy increases, the noise gradually increases, and the outline of the image becomes blurry. This occurs because the higher the energy of the incident of the gamma-rays is, the more gamma-rays that can penetrate through the shield area of the collimator, but not the holes, and finally reach the detector. This can enhance the noise and decrease the signal-to-noise ratio (SNR) of the image.

The artifacts are still visible in simplified respiratory images reconstructed using the MPA-MURA imaging system, particularly in the image of Tc-99m. Because the thickness of the collimator was optimized according to the 365-keV gamma rays from I-131, but not the 140-keV gamma rays from Tc-99m, the artifacts in the image of Tc-99m are considered slightly stronger than those of I-131.

In the MPA-MURA coded-aperture imaging, the effective aperture showed a strong relation with the energy and angle of the incident gamma rays. When the energy of the incident gamma rays shifts from a high point (365 keV) to a low point (140 keV), the effective sizes of the aperture will be changed. This means that the effective MPA-MURA of the coded-aperture collimator changes with the size of the distribution of the source and the mismatch between the thickness of the coded-aperture collimator and the energy of the incident gamma rays, which in turn can enhance the artifacts. In fact, when the thickness of the collimator was reduced by 1 mm, a clearer image, as shown in Fig. 6, was acquired. Therefore, the thickness of the collimator has a significant impact on the coded-aperture imaging. When the gamma-ray energy increases, the imaging quality decreases accordingly.

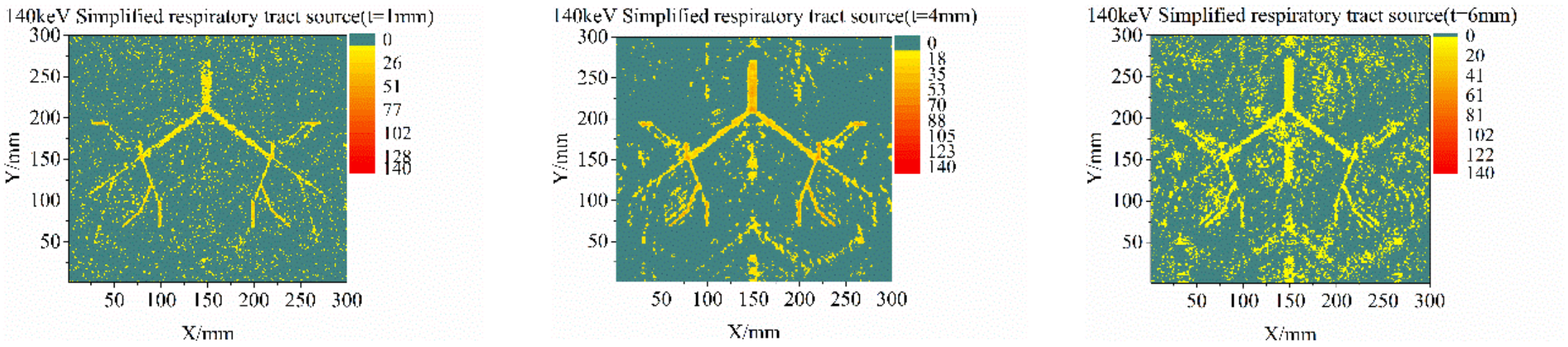

Fig. 6 Reconstruction of images of the simplified respiratory tract source of Tc-99m by the MPA-MURA coded-aperture collimator with different thicknesses (t = 1, 4, and 6 mm).

As shown in Figure 5, the images obtained by the parallel hole imaging have less noise than those obtained by the coded-aperture imaging, but the outlines of the former images are slightly more obvious than the latter ones at each energy point. The SNRs of the images presented in Fig. 5 are listed in Table 1.

From the Tc-99m to Cs-137 sources, the SNRs of the images obtained by parallel hole imaging are higher than those obtained by the coded-aperture imaging at each energy point. The SNRs of Co-60 images obtained by parallel-hole imaging were lower than those of Co-60 obtained by coded-aperture imaging.

The images of the I-131 source obtained by the MPA-MURA coded-aperture have the best SNR among the images obtained from the four nuclides in each type of distribution. The SNR of the images of the simplified respiratory tract source, obtained by the MPA-MURA coded-aperture imaging system, is the lowest of the three distributions.

Table 1 Signal-to-noise ratio and detection efficiency of images in Fig. 5

| **Radioactive source distribution** | **SNR（dB）** | | | | | | | |
|---|---|---|---|---|---|---|---|---|
| | $^{99m}Tc$(140 keV) | | $^{131}I$(365 keV) | | $^{137}Cs$(662 keV) | | $^{60}Co$(1332 keV) | |
| | PMA-MURA | Parallel hole | PMA-MURA | Parallel hole | PMA-MURA | Parallel hole | PMA-MURA | Parallel hole |
| A-shape | 4.49 | 19.94 | 6.04 | 17.41 | 4.92 | 12.26 | 3.90 | 2.17 |
| Ringed | 4.08 | 34.93 | 5.10 | 32.11 | 4.64 | 22.84 | 3.24 | 2.37 |
| Simplified respiratory tract | 3.33 | 13.98 | 3.98 | 13.19 | 3.37 | 9.23 | 3.11 | 1.80 |
| **Radioactive source distribution** | **Detection efficiency** | | | | | | | |
| | $^{99m}Tc$(140 keV) | | $^{131}I$(365 keV) | | $^{137}Cs$(662 keV) | | $^{60}Co$(1332 keV) | |
| | PMA-MURA | Parallel hole | PMA-MURA | Parallel hole | PMA-MURA | Parallel hole | PMA-MURA | Parallel hole |
| A-shape | $1.70 \times 10^{-3}$ | $4.03 \times 10^{-5}$ | $1.56 \times 10^{-3}$ | $4.99 \times 10^{-5}$ | $1.70 \times 10^{-3}$ | $9.17 \times 10^{-5}$ | $1.72 \times 10^{-3}$ | $2.50 \times 10^{-3}$ |
| Ringed | $1.82 \times 10^{-4}$ | $8.10 \times 10^{-6}$ | $1.66 \times 10^{-3}$ | $1.04 \times 10^{-5}$ | $1.83 \times 10^{-3}$ | $1.87 \times 10^{-5}$ | $1.85 \times 10^{-3}$ | $5.10 \times 10^{-3}$ |
| Simplified respiratory tract | $1.92 \times 10^{-3}$ | $5.60 \times 10^{-5}$ | $1.69 \times 10^{-3}$ | $6.79 \times 10^{-5}$ | $1.87 \times 10^{-3}$ | $1.28 \times 10^{-4}$ | $1.95 \times 10^{-3}$ | $3.10 \times 10^{-3}$ |

According to Table 1, with the increased radiant energy from the radiation source, the detection efficiency of the PMA-MURA coded-aperture imaging also increases accordingly, but the SNR of the images gradually decreases. This is because the transmission of high-energy rays leads to more rays penetrating the collimator and reaching the detector, contributing to the total count and increasing the amount of noise. According to Table 1, the detection efficiency of the PMA-MURA coded aperture imaging was two orders of magnitude higher than that of parallel hole collimator imaging, except for Co-60 images. The energy of gamma rays emitted by the Co-60 sources was too high for the two collimators (4 mm) and could be penetrated by a large amount of 1332 keV gamma rays. Therefore, for low-activity internal pollution imaging, PMA-MURA coded aperture imaging with a high detection efficiency and considerable spatial resolution makes it a good option for internal pollution imaging measurements.

## 7. CONCLUSION

Internal contamination imaging using a PMA-MURA coded aperture collimator was simulated in the lungs. The MLEM algorithm was used to reconstruct the radioactive source image. The noise and artifacts in the reconstructed images were reduced using the complementary imaging method. The point source, A-shaped source, ringed source, and simplified respiratory tract source placed in the lung model were imaged. The spatial resolution (FWHM) of the I-131 (365 keV) point source obtained by the PMA-MURA coded-aperture collimator was 2.51 mm. Although the SNR of the PMA-MURA coded-aperture imaging is much lower than that of the parallel hole collimator imaging, a clear distribution of radioactive nuclides (Tc-99m, I-131, Cs-137, and Co-60) can still be obtained in the lungs. The spatial resolution and noise of the images are related to the specifications of the PMA-MURA coded-aperture collimator when the imaging geometry is fixed. It is important to note that the detection efficiency of PMA-MURA coded aperture imaging is two orders of magnitude higher than that of parallel hole imaging. Considering the low activity level of the internal lung contamination caused by nuclear accidents, PMA-MURA coded-aperture imaging has a significant potential for the development of lung contamination imaging.

**Acknowledgements**

This study was supported by the Ministry of Science and Technology, People's Republic of China (Contract No. 2012YQ180118), which provided the major scientific instruments and equipment development. This study was also supported by the National Natural Science Foundation of China (No. 41874121) and the Sichuan science and technology program (No. 2018JY0181).

---